\documentclass[letterpaper]{aa}
\usepackage{graphicx}
\usepackage{natbib}
\bibpunct{(}{)}{;}{a}{}{,}
\usepackage{txfonts}
\begin{document}
   \title{Precise radial velocities of giant stars \thanks{Based on observations taken at University of California Observatories / Lick Observatory.}}

   \subtitle{I. Stable stars}

   \author{S. Hekker  \inst{1}
    \and S. Reffert \inst{1}
     \and A. Quirrenbach \inst{1}
      \and D.S. Mitchell \inst{2}
       \and D.A. Fischer \inst{3}
        \and G.W. Marcy \inst{4}
	 \and R.P. Butler \inst{5}}

   \offprints{S. Hekker, \\
                    email: saskia@strw.leidenuniv.nl}

   \institute{Leiden Observatory, Leiden University, P.O. Box 9513, 2300 RA Leiden, Netherlands
                  \and
                    California Polytechnic State University, San Luis Obispo, CA 93407, USA
                    \and
                    Department of Physics and Astronomy, San Francisco State University, 1600 Holloway, San Francisco, CA 94132, USA
                    \and
                    Department of Astronomy, University of California at Berkeley, 601 Campbell Hall, Berkeley, CA 94720, USA
                    \and
                   Department of Terrestrial Magnetism, Carnegie Institution of Washington, 5241 Broad Branch Road, NW, Washington, DC 20015-1305, USA     }

   \date{Received 1 February 2006  / Accepted 18 April 2006}


   \abstract{Future astrometric missions such as SIM PlanetQuest need very stable reference stars. K giants have large luminosities, which place them at large distances and thus the jitter of their photocenters by companions is relatively small. Therefore K giants would be best suited as references. To confirm this observationally a radial velocity survey is performed to quantify the level of intrinsic variability in K giants.}
   {From this radial velocity survey we present 34 K giants with an observed standard deviation of the radial velocity of less than 20 m/s. These stars are considered ``stable'' and can be used as radial velocity standards.}
   {The radial velocity survey contains 179 K giants. All K giants have a declination between $-30\degr$ and $+65\degr$ and visual magnitude of $3-6$ mag. The Coud\'e Auxiliary Telescope (CAT) at UCO/Lick Observatory is used to obtain radial velocities with an accuracy of $5-8$ m/s. The number of epochs for the 34 stable stars ranges from 11 to 28 with a total timespan of the observations between 1800 and a little over 2200 days. }
   {The observational results of the 34 ``stable'' stars are shown together with a discussion about their position in the $M_{V}$ vs. $B-V$ diagram and some conclusions concerning the radial velocity variability of K giants. These results are in agreement with the theoretical predictions. K giants in a certain range of the $M_{V}$ vs. $B-V$ diagram are suitable reference stars.}
   {}
   
   
  \authorrunning{S. Hekker et al.}
   \maketitle
%

\section{Introduction}
To perform high precision astrometric observations very stable reference stars are needed. In preparation of the Space Interferometry Mission (SIM PlanetQuest), \citet{frink2001} investigated which type of stars would be best suited as reference stars. Although known to be photospherically active, K giants appeared to be the best choice, mainly because of their large distances, brightness and sky coverage. To quantify the photospheric activity observationally, a radial velocity survey was started to measure the level of intrinsic radial velocity variability in K giant stars.

For about a decade, well known techniques have been used to perform very accurate radial velocity observations, up to a few m/s \citep[see e.g.][]{marbut2000,queloz2001} and with HARPS (High Accuracy Radial velocity Planet Searcher on the 3.6~m telescope, La Silla Observatory, ESO Chile) even to 1 m/s \citep{pepe2003}. Most extrasolar planets known so far have been discovered around main sequence stars using radial velocity observations. Like main sequence stars, K giant spectra contain a large number of narrow spectral lines and accurate radial velocity variations can also be obtained for these stars.

In this paper we present results for 34 K giants, from the above-mentioned survey, with an observed standard deviation of the radial velocity of less than 20 m/s. These stars are considered stable and can be used as radial-velocity standards.

In general the possibility of accurate radial velocity observations makes it possible and necessary to select radial velocity standards with smaller radial velocity variations. The IAU standard stars \citep{pearce1955} and the suggested extensions by \citet{heard1968} and \citet{evans1968} for the northern and southern sky respectively do not yet have an accuracy of a few m/s. More recently \citet{kharchenko2004} selected 3967 stars from their ``Catalog of radial velocities of galactic stars with high precision astrometric data (CRVAD)'' \citep[based on][]{barbier2000} as radial velocity standard candidates. Furthermore \citet{udry1999a,udry1999b} present a list of CORAVEL radial-velocity standard stars, and a list with proposed high-precision radial-velocity standards, respectively. The stars presented in this paper are in addition to the already known radial velocity standard stars.

The paper is organized as follows. In Section 2 the observations are described, followed in Section 3 by the results for the individual stars. Section 4 contains a discussion and some conclusions concerning the radial velocity variability of K giants.
   

\section{Observations}

The sample of 179 K giants has been selected from the Hipparcos catalog \citep{esa1997} based on the criteria described in \citet{frink2001}. They are all brighter than 6~mag, presumably single, and have masses ranging from about 1 to 3 solar masses. In Table~\ref{tab1} properties of the 34 stable stars are listed.

Our ongoing K giant radial velocity survey started in June 1999, with the Coud\'e Auxiliary Telescope (CAT) in conjunction with the Hamilton high resolution (R=60\,000) echelle spectrograph. An iodine cell is placed in the light path. With integration times of up to thirty minutes for the faintest stars we reach a signal to noise ratio of about $80-100$, yielding a radial velocity precision of $5-8$ m/s. This is adequate for our survey and hence no attempt has been made to reach the 3 m/s accuracy which is in principle possible with this setup \citep{butler1996}. The pipeline described by \citet{butler1996} is used. A template iodine spectrum and a template spectrum of the target star obtained without an iodine cell in the lightpath are used to model the stellar observations obtained with an iodine cell in the lightpath. The Doppler shift is a free parameter in this model and determined as the shift of the template stellar spectrum to obtain the best model for the observed spectra. With this method the radial velocity itself is not measured. Only the change in the radial velocity with respect to the stellar template is obtained with a precision of a few m/s. The mean radial velocities of the stars are known with an accuracy of the order of a few tenths of km/s from for instance \citet{famaey2005} and \citet{barbier2000}. The radial velocities from \citet{famaey2005} were obtained with the CORAVEL spectrovelocimeter mounted on the swiss 1~m-telescope at the Observatoire Haute Provence, France. These are more accurate than the ones from \citet{barbier2000}, but not available for all stars in our sample. The latter is an extension of the WEB Catalog of Radial Velocities \citep{duflot1995}.

\section{Results}
34 stars out of the sample of 179 stars have an observed standard deviation of the radial velocity of less than 20 m/s. The exact value of this threshold is somewhat arbitrary. It is set by a visual inspection of the radial velocity variations observed in our sample. Stars without systematic radial velocity variations or trends all happen to have an observed standard deviation of the radial velocity of less than 20 m/s. Furthermore, selecting reference stars for SIM PlanetQuest with radial velocity variations smaller than 20 m/s would result in an acceptable $3.6\%$ contamination of the reference star grid with binary stars \citep{frink2001}. Plots of the radial velocity variation are shown in Figure~\ref{constants0}. The numbers in the upper right corner of each frame denote the observed standard deviation ($\sigma_{std}$, upper number) and the mean error ($\sigma_{me}$, lower number) of the radial velocity observations. The latter is derived from the rms scatter of hundreds of individual ``chunks'' of the spectrum, typically $2\AA$. In Table~\ref{tab2}, for each of the 34 stable stars the mean error, number of observations, and timespan of the observations in this survey are listed together with the observed standard deviation, intrinsic standard deviation (which is obtained by quadratically subtracting the mean error from the total observed radial velocity scatter), and the reduced $\chi^{2}$. Furthermore a flag is set to K for stars also present among the 3967 candidate standards presented by \citet{kharchenko2004}. In case the flag is set to N, there are not enough observations in \citet{kharchenko2004} to make it a radial velocity standard candidate, but all other parameters do match their stability criteria.

All other stars in the sample show radial velocity variations larger than 20 m/s. Around one star a substellar companion has been discovered \citep[$\iota$ Draconis,][]{frink2002}. The highly non-sinusoidal radial velocity variation observed for this star can only be induced by a companion with high eccentricity and not by stellar activity. This star also shows a long-term trend indicating a third component in the system. About 23 spectroscopic binaries are present. Some are already known in the literature, but some were not observed before and will be presented in a forthcoming paper \citep{reffert2006}. Furthermore, about 35 stars with sinusoidal periodic variations are present \citep{hekker2006}, among which four show an additional long trend indicating a binary in a wide orbit. The nature of these sinusoidal periodic variations is under investigation. For at least four stars there are strong arguments that the presence of nearly sinusoidal variations of the radial velocity are most likely caused by substellar companions \citep{mitchell2006}. Two of the stars with very large radial velocity variations of several km/s appear to be supergiants. A summary of the whole program is presented in \citet{mitchell2006}.

\begin{table*}[!]
\begin{minipage}[t]{17cm}
\caption{Properties of the stable stars:  right ascension ($ra$) in ``hh:mm:ss" and declination ($dec$) in ``dd:mm:ss", both J2000.0, apparent magnitude ($m_{V}$) and absolute magnitude ($M_{V}$) in the V band, parallax ($plx$) in mas, $B-V$ color, (rather uncertain) mass obtained with the method described by \citet{allende1999} in M$_{\sun}$, the spectral type ($SP$) and the radial velocity $RV$ in km/s from \citet{famaey2005} and \citet{barbier2000}, respectively. The latter catalog does not give errors in the radial velocity for each star. }
   \label{tab1}
    \centering
    \renewcommand{\footnoterule}{}
    \begin{tabular}{llcccrrcclrr}
    \hline\hline
    HIP & HD & $ra$ & $dec$ & $m_{V}$\footnote{The Hipparcos and Tycho Catalogues \citep{esa1997}}  & $M_{V}$ & $plx^{a}$ & $B-V^{a}$  & mass & $SP^{a}$ & $RV$ \footnote{Radial velocities for 6691 K and M giants \citep{famaey2005} }& $RV$\footnote{General Catalog of mean radial velocities \citep{barbier2000}} \\
     & & hh:mm:ss & dd:mm:ss & mag & mag & mas & mag  & M$_{\sun}$ & & [km/s] & [km/s] \\
    \hline
    \object{HIP4906} & \object{HD6186} & 01 02 56.6 & $+$07 53 25 & 4.27 & 0.44 & 17.14 & 0.952 & 2.27 &K0III & 7.47 $\pm$ 0.20 & 7.50 $\pm$ 0.2\\
    \object{HIP13701} & \object{HD18322} & 02 56 25.7 & $-$08 53 53 & 3.89 & 0.83 & 24.49 & 1.088 & 1.38 &K1III-IV & &  $-$20.30 ~~~~~~~~~     \\
    \object{HIP14838} & \object{HD19787} & 03 11 37.8 & $+$19 43 36 & 4.35 & 0.79 & 19.44 & 1.033 & 1.91 & K2III & 23.05 $\pm$ 0.20 & 23.90 $\pm$ 0.4\\
    \object{HIP19388} & \object{HD26162} & 04 09 10.0 & $+$19 36 33 & 5.51 & 0.76 & 11.21 & 1.077 & 1.39 & K2III & 24.75 $\pm$ 0.02 & 24.80 ~~~~~~~~~\\
    \object{HIP21248} & \object{HD29085} & 04 33 30.6 & $-$29 45 59 & 4.49 & 1.58 & 26.22 & 0.972 & 1.98 & K0III & &  20.60 ~~~~~~~~~\\
    \object{HIP22860} & \object{HD31414} & 04 55 06.8 & $-$16 44 26& 5.71 & $-$0.11 & 6.85 & 0.953 & 3.01 &K0II & & 9.80 ~~~~~~~~~\\
    \object{HIP33914} & \object{HD52556} & 07 02 17.5 & $+$15 20 10 & 5.78 & $-$0.70 & 5.06 & 1.140 & 3.06 &K1III & $-$12.85 $\pm$ 0.20 &$-$13.50 ~~~~~~~~~\\
    \object{HIP36848} & \object{HD60666} & 07 34 34.8 & $-$27 00 44 & 5.78 & 0.87 & 10.41 & 1.045 & 1.71 & K1III & & $-$6.20 $\pm$ 0.3\\
    \object{HIP37447} & \object{HD61935} & 07 41 14.8 & $-$ 09 33 04 &  3.94 & 0.71 & 22.61& 1.022  & 1.94 & K0III & &  10.50 ~~~~~~~~~\\
    \object{HIP38375} & \object{HD64152} & 07 51 43.0 & $-$21 10 25 & 5.62 & 1.00 & 11.90 & 0.956 & 2.50 & K0III & &  31.90 ~~~~~~~~~\\
    \object{HIP43923} & \object{HD76291} & 08 56 50.0 & $+$45 37 54 & 5.72 & 1.48 & 14.21 & 1.125 & 1.30 & K1IV & 53.28 $\pm$ 0.30 & 58.40 $\pm$ 0.3\\
    \object{HIP48455} & \object{HD85503} & 09 52 45.8 & $+$26 00 25 & 3.88 & 0.83 & 24.52 & 1.222 & 0.59 & K0III & 13.63 $\pm$ 0.07 & 14.10 $\pm$ 0.3\\
    \object{HIP53316} & \object{HD94481} & 10 54 17.8 & $-$13 45 29 & 5.65 & 0.16 & 7.97 & 0.832 & 2.89 & K0III & & 5.40 ~~~~~~~~~\\
    \object{HIP58181} & \object{HD103605} & 11 55 58.4 & $+$56 35 55 & 5.83 & 0.90 & 10.34 & 1.101 & 1.45 & K1III & 16.91 $\pm$ 0.16 & 14.70 $\pm$ 1.2\\
    \object{HIP59847} & \object{HD106714} & 12 16 20.5 & $+$23 56 43 & 4.93 & 0.52 & 13.12 & 0.957 & 2.27 & K0III & $-$27.89 $\pm$ 0.13 & $-$27.20 $\pm$ 0.5\\
    \object{HIP60742} & \object{HD108381} & 12 26 56.3 & $+$28 16 06 & 4.35 & 0.76 & 19.18 & 1.128 & 1.66 & K2III & 3.38 $\pm$ 0.11 & 4.70 $\pm$ 0.3\\
    \object{HIP68895} & \object{HD123123} & 14 06 22.3 & $-$26 40 57 & 3.25 & 0.79 & 32.17 & 1.091 & 1.76 & K2III & & 27.20 $\pm$ 0.5\\
    \object{HIP74239} & \object{HD134373} & 15 10 18.6 & $-$26 19 57 & 5.75 & 0.05 & 7.25 & 1.045 & 2.78 & K0III & & $-$33.10 $\pm$ 0.3\\
    \object{HIP75944} & \object{HD138137} & 15 30 40.4 & $-$16 36 34 & 5.82 & $-$0.37 & 5.78 & 1.056 & 2.94 & K0III & &  $-$1.70 ~~~~~~~~~\\
    \object{HIP78132} & \object{HD142980} & 15 57 14.6 & $+$14 24 52 & 5.54 & 1.33 & 14.36 & 1.141 & 1.19 & K1IV & $-$70.98 $\pm$ 0.17 & $-$68.30 ~~~~~~~~~\\
    \object{HIP78442} & \object{HD143553} & 16 00 61.1 & $+$04 25 39 & 5.82 & 1.49 & 13.62 & 1.003 & 1.93 & K0III & $-$7.85 $\pm$ 0.22 & $-$4.10 ~~~~~~~~~\\
    \object{HIP83000} & \object{HD153210} & 16 57 10.1 & $+$09 22 30 & 3.19 & 1.09 & 37.99 & 1.160 & 0.78 & K2III & $-$55.86 $\pm$ 0.19 & $-$54.40 $\pm$ 1.3\\
    \object{HIP88684} & \object{HD165438} & 18 06 15.2 & $-$04 45 05 & 5.74 & 3.02 & 28.61 & 0.968 & 1.35 & K1IV & &  $-$18.90 ~~~~~~~~~\\
    \object{HIP89962} & \object{HD168723} & 18 21 18.6 & $-$02 53 56 & 3.23 & 1.84 & 52.81 & 0.941 & 1.96 & K0III-IV & &  8.90 $\pm$ 0.7\\
    \object{HIP90496} & \object{HD169916} & 18 27 58.2 & $-$25 25 18 & 2.82 & 0.95 & 42.20 & 1.025 & 1.88 & K1III & &  $-$43.20 $\pm$ 0.7\\
    \object{HIP93085} & \object{HD175775} & 18 57 43.8 & $-$21 06 24 & 3.52 & $-$1.77 & 8.76 & 1.151 & 4.58 & K0II-III & & $-$20.10 $\pm$ 0.6\\
    \object{HIP94779} & \object{HD181276} & 19 17 06.2 & $+$53 22 06 & 3.80 & 0.91 & 26.48 & 0.950 & 2.87 & K0III & $-$29.00 $\pm$ 0.30 & $-$29.20 $\pm$ 0.6\\
    \object{HIP96229} & \object{HD184406} & 19 34 05.4 & $+$07 22 44 & 4.45 & 1.80 & 29.50 & 1.176 & 0.92 & K3III & $-$24.73 $\pm$ 0.13 & $-$23.90 $\pm$ 0.6\\
    \object{HIP96459} & \object{HD185351} & 19 36 38.0 & $+$44 41 42 & 5.17 & 2.13 & 24.64 & 0.928 & 1.82 & K0III & $-$5.91 $\pm$ 0.11 & $-$5.20 $\pm$ 1.0\\
    \object{HIP102422} & \object{HD198149} & 20 45 17.4 & $+$61 50 20 & 3.41 & 2.63 & 69.73 & 0.912 & 1.64 & K0IV & $-$87.55 $\pm$ 0.11 & $-$87.90 $\pm$ 0.6\\
    \object{HIP106039} & \object{HD204381} & 21 28 43.4 & $-$21 48 26 & 4.50 & 0.80 & 18.18 & 0.889 & 3.46 & K0III & &  $-$20.80 $\pm$ 0.8\\
    \object{HIP112724} & \object{HD216228} & 22 49 40.8 & $+$66 12 01 & 3.50 & 0.76 & 28.27 & 1.053 & 1.61 & K0III & $-$12.59 $\pm$ 0.20 & $-$14.20 $\pm$ 0.7\\
    \object{HIP115438} & \object{HD220321} & 23 22 58.2 & $-$20 06 02 & 3.96 & 0.48 & 20.14 & 1.082 & 2.06 & K0III & & $-$6.10 $\pm$ 0.4\\
    \object{HIP115830} & \object{HD220954} & 23 27 58.1 & $+$06 22 44 & 4.27 & 0.83 & 20.54 & 1.062 & 1.54 & K1III & 6.05 $\pm$ 0.19 & 6.50 $\pm$ 1.8\\
    \hline
   \end{tabular} 
   \end{minipage}
 \end{table*}
 
\begin{table*}
\centering
\caption{Observational results of the 34 stable stars: the mean error of the individual observations for each star in m/s, the number of observations ($N$), the time span of the observations in days, the standard deviation of the radial velocity ($\sigma$) in m/s, the intrinsic scatter ($\sigma_{int}$) obtained by quadratically subtracting the mean error from the total observed radial velocity scatter, the reduced $\chi^2$ and a flag. This flag is set to K for all stars also present among the 3967 candidate standards presented by \citet{kharchenko2004}. The flag is set to N for the stars with less than 4 observations in \citet{kharchenko2004} which do otherwise match their criteria. If the flag is blank a photometric variability flag is present in the Tycho~1 catalog \citep{esa1997} and these stars are therefore not included in the \citet{kharchenko2004} candidate standard star catalog. More details about the selection method used by \citet{kharchenko2004} are described in the text.}
    \label{tab2}
     \begin{tabular}{llrrrrrrc}
    \hline\hline
    HIP & HD & mean error & $N$ & timespan & $\sigma$ & $\sigma_{int}$ & $\chi^2_{r}$ & flag\\
    & & [m/s] & & [days] & [m/s] & [m/s] & &\\
    \hline
    \object{HIP4906} & \object{HD6186} & 5.2  & 19 & 1837 & 15.9 &15.0 & 9.2 &\\
    \object{HIP13701} & \object{HD18322} & 5.0 & 28 & 1806 & 14.9 &14.0 & 8.6 &\\
    \object{HIP14838} & \object{HD19787} & 4.7 & 18 & 1803 & 12.3 & 11.4 & 8.3 &\\
    \object{HIP19388} & \object{HD26162} & 5.7 & 18 & 2222 & 16.2 & 15.2 & 7.6 &K\\
    \object{HIP21248} & \object{HD29085} & 6.1 & 16 & 1832 & 15.7 & 14.5 & 7.5 &\\
    \object{HIP22860} & \object{HD31414} & 7.8 & 19 & 2216 & 12.2 & 9.4 & 2.4 & N\\
    \object{HIP33914} & \object{HD52556} & 6.8 & 20 & 1877 & 18.2 & 16.9 & 6.6 & K\\
    \object{HIP36848} & \object{HD60666} & 8.4 & 11 & 1874 & 16.6 & 14.3 & 3.8 & K\\
    \object{HIP37447} & \object{HD61935} & 5.0 & 14 & 1839 & 15.5 & 14.7 & 9.6 & \\
    \object{HIP38375} & \object{HD64152} & 6.9 & 11 & 1897 & 15.9 & 14.3 & 5.8 & N\\
    \object{HIP43923} & \object{HD76291} & 7.1 & 13 & 1718 & 14.4 & 12.5 & 4.1 & K\\
    \object{HIP48455} & \object{HD85503} & 4.3 & 14 & 1458 & 20.0 & 19.5 & 23.6 &\\
    \object{HIP53316} & \object{HD94481} & 10.3 & 11 & 1814 & 16.1 & 12.4 & 2.3 & N\\
    \object{HIP58181} & \object{HD103605} & 6.5 & 13 & 1951 & 12.7 & 10.9 & 4.2 & K\\
    \object{HIP59847} & \object{HD106714} & 7.5 & 14 & 1882 & 11.1 & 8.2 & 2.6 & K\\
    \object{HIP60742} & \object{HD108381} & 5.4 & 16 & 1545 & 12.4 & 11.2 & 5.8 &\\
    \object{HIP68895} & \object{HD123123} & 5.2 & 16 & 1688 & 19.8 & 19.1 & 14.9 &\\
    \object{HIP74239} & \object{HD134373} & 7.7 & 15 & 1935 & 15.3 & 13.2 & 4.5 & K\\
    \object{HIP75944} & \object{HD138137} & 7.4 & 16 & 1933 & 16.1 & 14.3 & 3.3 & N\\
    \object{HIP78132} & \object{HD142980} & 5.6 & 16 & 1932 & 19.0 & 18.2 & 12.0 & K\\
    \object{HIP78442} & \object{HD143553} & 5.7 & 16 & 1933 & 13.2 & 11.9 & 5.4 & N\\
    \object{HIP83000} & \object{HD153210} & 4.5 & 17 & 1872 & 16.7 & 16.1 & 13.4 &\\
    \object{HIP88684} & \object{HD165438} & 5.7 & 20 & 2200 & 17.9 & 17.0 & 10.8 & N\\
    \object{HIP89962} & \object{HD168723} & 5.2 & 18 & 1876 & 13.4 & 12.4 & 7.0 &\\
    \object{HIP90496} & \object{HD169916} & 5.0 & 18 & 1879 & 12.8 & 11.8 & 7.0 &\\
    \object{HIP93085} & \object{HD175775} & 5.6 & 17 & 1899 & 17.0 & 16.1 & 8.4 &\\
    \object{HIP94779} & \object{HD181276} & 4.8 & 18 & 1835 & 9.9 & 8.7 & 4.7 &\\
    \object{HIP96229} & \object{HD184406} & 4.6 & 19 & 1843 & 17.0 & 16.4 & 14.3 &\\
    \object{HIP96459} & \object{HD185351} & 5.6 & 25 & 2233 & 9.7 & 7.9 & 3.4 & K\\
    \object{HIP102422} & \object{HD198149} & 5.1 & 20 & 1874 & 10.4 & 9.1 & 4.8 &\\
    \object{HIP106039} & \object{HD204381} & 5.9 & 18 & 1908 & 8.8 & 6.5 & 2.1 &\\
    \object{HIP112724} & \object{HD216228} & 4.4 & 20 & 1839 & 12.1 & 11.3 & 8.2 & K\\
    \object{HIP115438} & \object{HD220321} & 5.2 & 21 & 1836 & 19.4 & 18.7 & 14.7 & K\\
    \object{HIP115830} & \object{HD220954} & 4.7 & 18 & 1904 & 12.4 & 11.5 & 7.2 & K\\
    \hline
   \end{tabular} 
 \end{table*}
 
\begin{figure*}
\centering
\includegraphics[width=17cm]{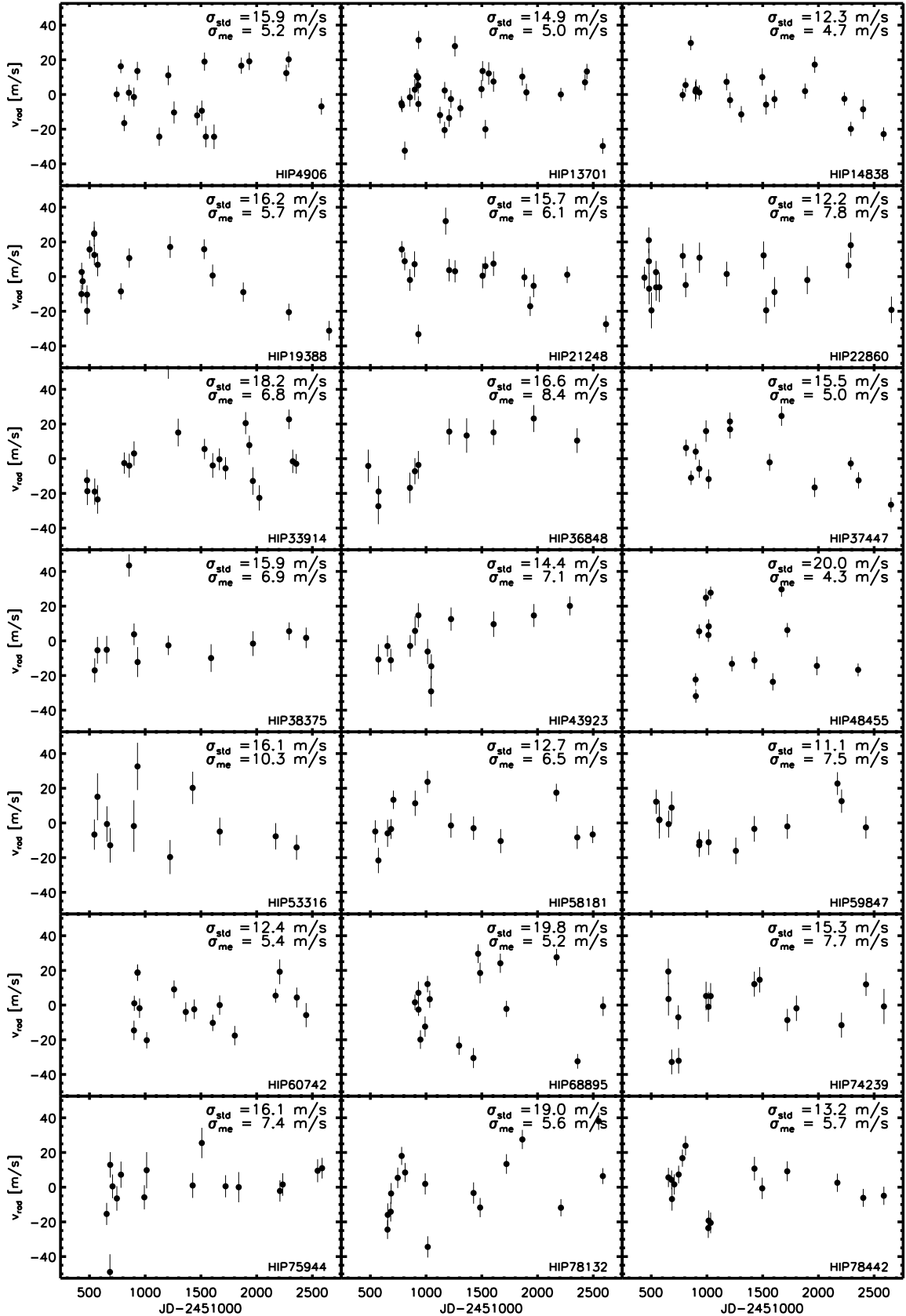}
\caption{ Radial velocity variations with an arbitrary zero point as a function of Julian date for the first 21 stars. The numbers in the upper right corner of each frame are the observed standard deviation ($\sigma_{std}$, upper number) and the mean error ($\sigma_{me}$, lower number) of the radial velocity observations. The Hipparcos catalog number is plotted in the lower right corner of each frame.}
\label{constants0}
\end{figure*}

\addtocounter{figure}{-1}
\begin{figure*}
  \centering
   \includegraphics[width=17cm]{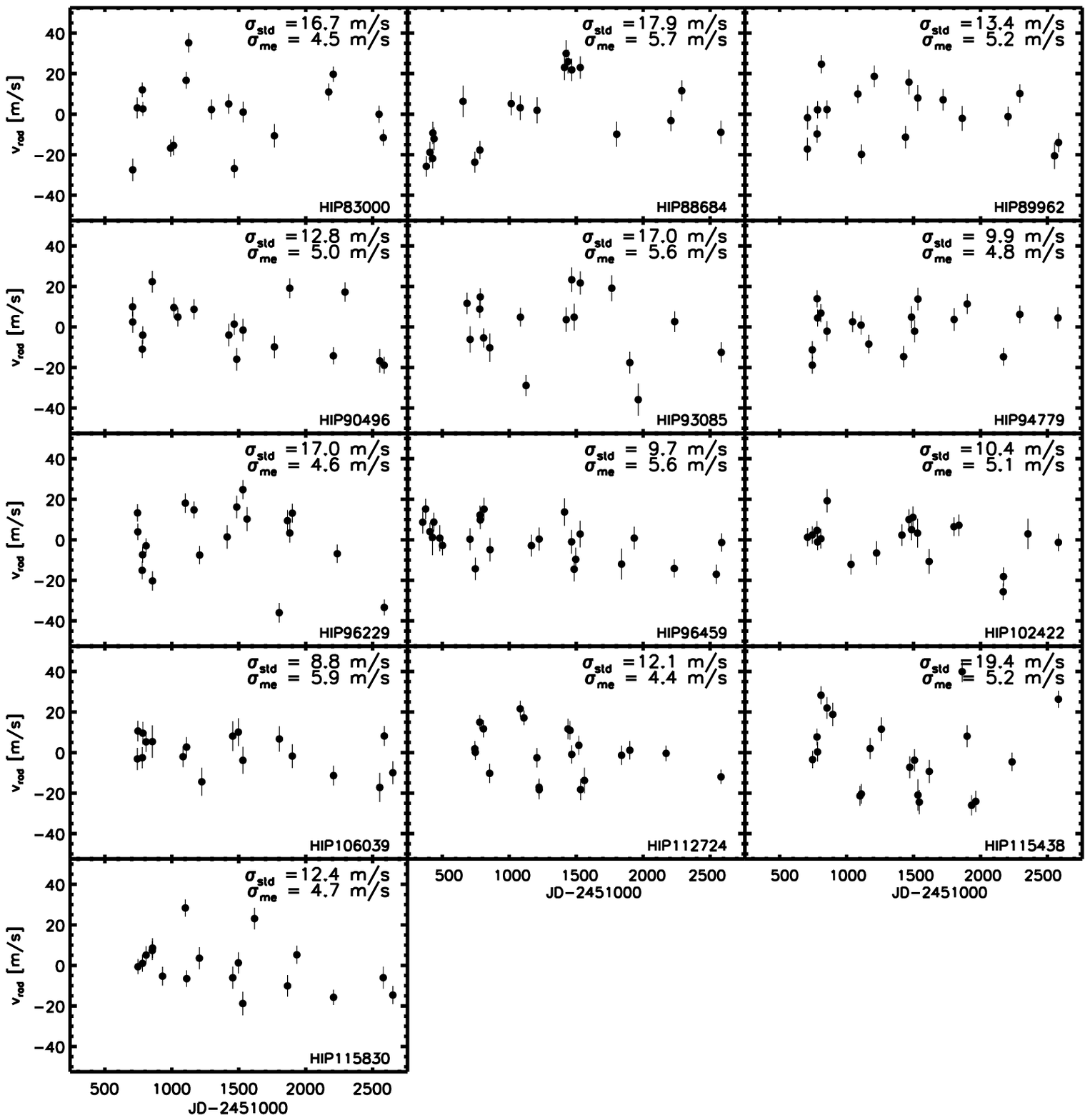}
      \caption{ continued for stars 22-34.}
       \label{constants1}
   \end{figure*}

\begin{figure}
 \resizebox{\hsize}{!}{\includegraphics{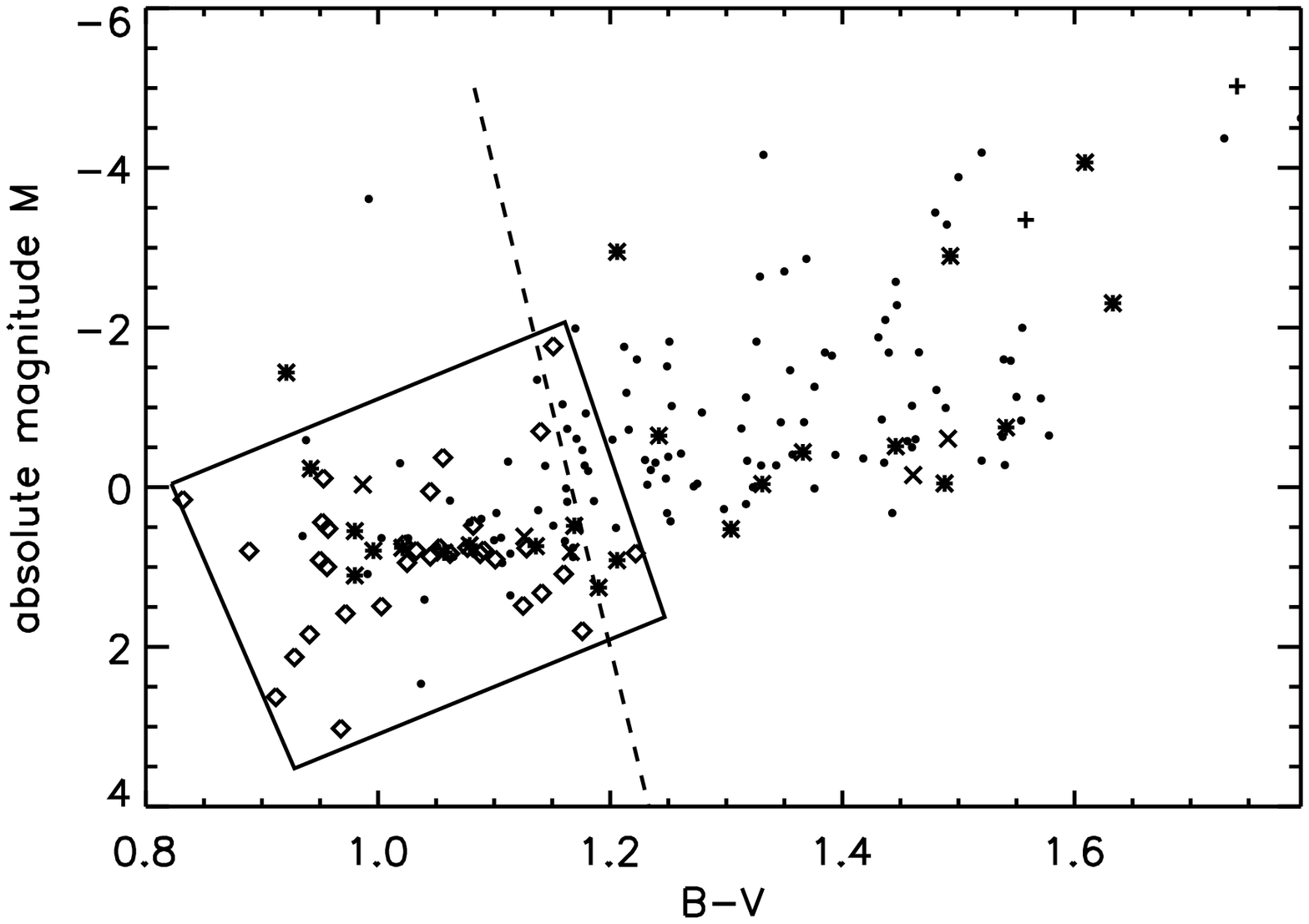}}
      \caption{$M_{V}$ vs. $B-V$ diagram with all 179 stars in the survey. The diamonds represent the stable stars, the asterisks represent the binaries, the crosses the variable stars with a long trend, the plus-signs are two supergiants with large, random radial velocity variations, and the dots are all other stars. The box is drawn around the stable stars and contains 11 binaries, 3 variable stars with a long trend, and 73 stars among which the 34 ``stable'' stars. The dashed line indicates the coronal dividing line (CDL) \citep{haisch1999}.}
         \label{HRdiagram}
   \end{figure}

        \begin{figure}
      \resizebox{\hsize}{!}{\includegraphics{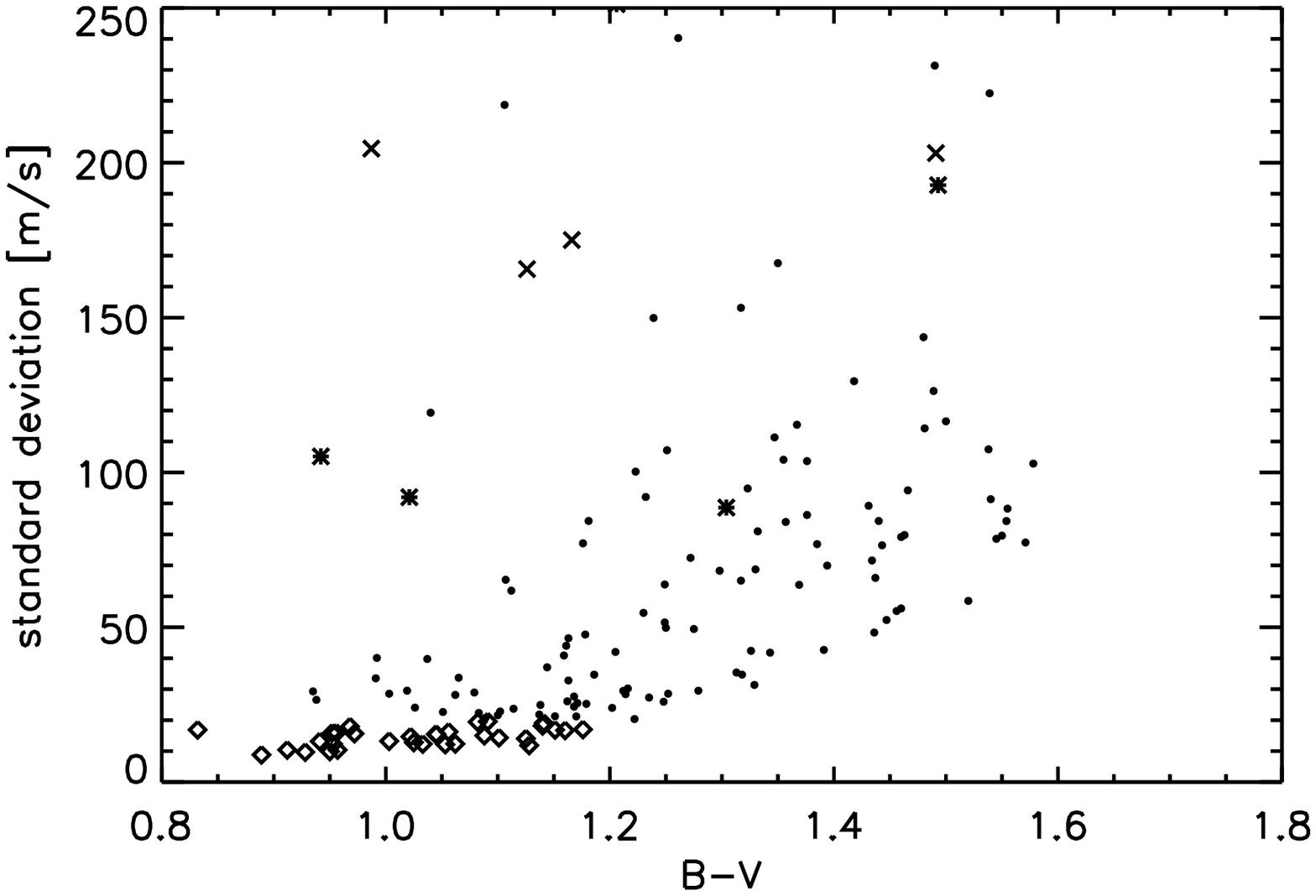}}
            \caption{Standard deviation of the radial velocity of the stars from the survey plotted as a function of $B-V$. Most stars with $B-V <1.2$ show smaller variations in the radial velocity than the ones with $B-V >1.2$. The symbols are the same as in Figure~\ref{HRdiagram}. The stars with a standard deviation larger than 250 m/s, are not shown.}
         \label{stddev}
   \end{figure}
   
  \begin{figure}
   \resizebox{\hsize}{!}{\includegraphics{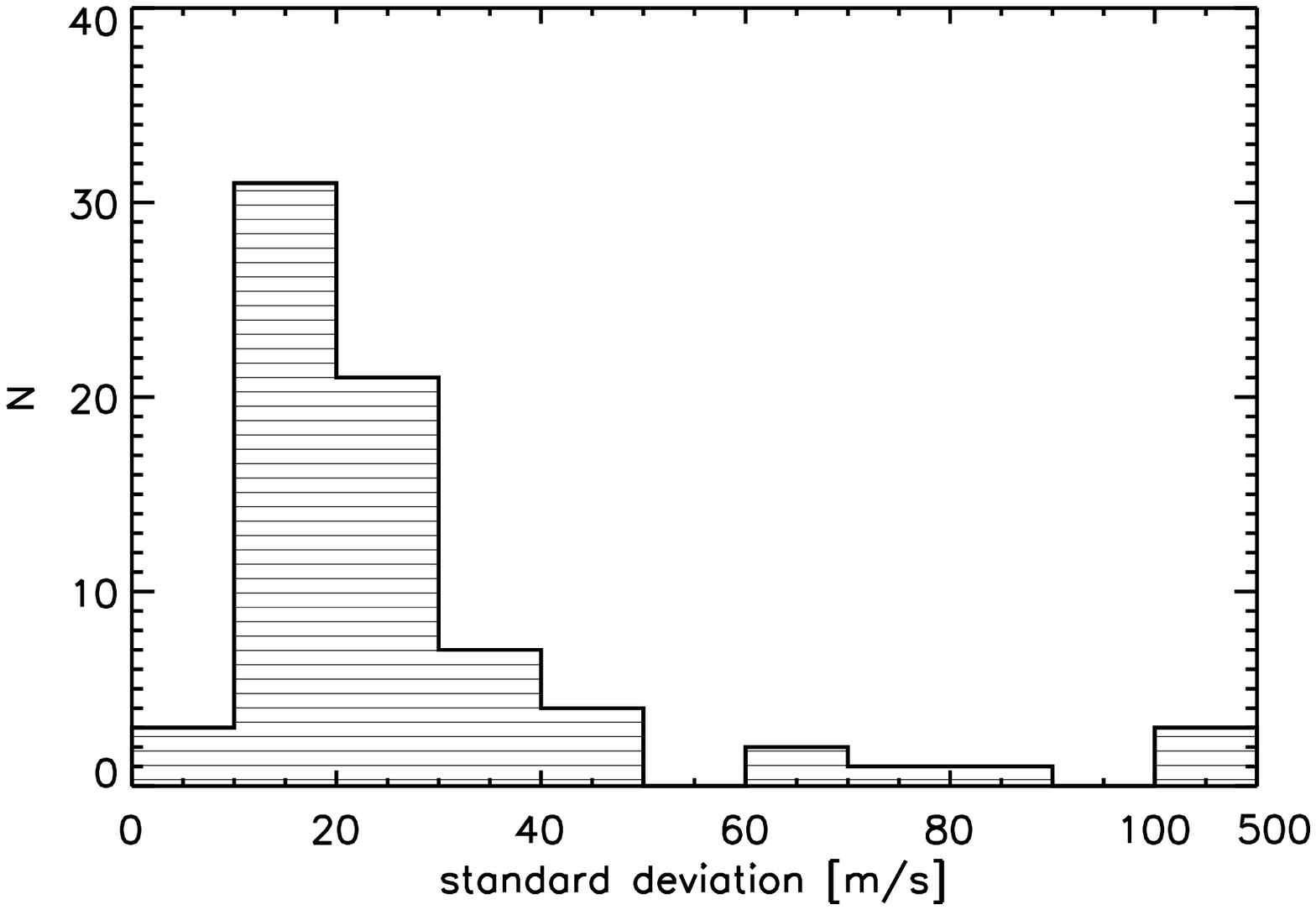}}
     \caption{Histogram of the observed standard deviations (including the contribution of the radial velocity errors) for the stars in the box in the $M_{V}$ vs. $B-V$ diagram from Figure~\ref{HRdiagram}. Binaries and the variable stars with a long trend are excluded. The stars in the highest bin have a standard deviation between 100 m/s and 500 m/s}
    \label{stddevhisto}
   \end{figure}

\section{Discussion \& Conclusions}
To obtain more information on the type of stars that appear to be stable, all 179 stars from this survey are plotted in an $M_{V}$ vs. $B-V$ diagram, see Figure~\ref{HRdiagram}. A box is drawn around the stable stars. The box contains 11 binaries, three variable stars with long trends indicating that they are binaries, and 73 other stars among which the 34 stable stars. The binaries and variable stars with a long trend are excluded in the further discussion.

The stable stars are not homogeneously distributed in the $M_{V}$ vs. $B-V$ diagram but they all have a $B-V$ color less than 1.2. This is shown in Figure~\ref{stddev}, where the standard deviation of the radial velocity is plotted as a function of $B-V$ color. The majority of the bluer stars show smaller variations in the radial velocity than the redder ones. This increase of the radial velocity variability with $B-V$ color, first described by \citet{frink2001}, is consistent with similar trends of photometry and radial velocity variability with spectral type (\citet{hatzes1998}, \citet{larson1999}, \citet{nidever2002}). These results are also in good agreement with the results by \citet{henry2000}. They obtained photometric observations of 187 G, K and M0 field giants and show that  ``stable'' giants with a short-term standard deviation less than 0.0020 mag have a $B-V$ less than 1.35. They note that nearly all stable giants are on the left side of the coronal dividing line (CDL). The CDL separates the giants with hot coronae on the left from giants with cool, massive winds on the right \citep{linskyhaisch1979,haisch1999}. 

\subsection{Statistics}
In Figure~\ref{stddevhisto}, a histogram of the standard deviation of the radial velocity from the stars in the box (Figure~\ref{HRdiagram}) is shown. Nearly half of the stars have a radial velocity with a standard deviation less than 20 m/s, 90~\% less than 50 m/s and 96~\% less than 100 m/s. By selecting stars from the region in the color-magnitude diagram outlined by this box, it is thus possible to construct samples of K giants with small radial velocity variations.

\subsection{Variability}
For each star the standard deviation of the observed radial velocities, although small, is significantly larger than the measurement errors, which implies that the stars show low-level radial velocity variations. To quantify this a $\chi^{2}$ test is performed to obtain the probability of the reality of this variability. The reduced $\chi^{2}$ values are listed in Table~\ref{tab2}. 31 of the 34 stars have a $>99.9\%$ probability of variability, for the other 3 the probability is $>99\%$. 
The presence of variability is consistent with the findings of \citet{barban2004} who detected solar-like oscillations in two red giant stars. One of these stars, \object{HIP89962}/\object{HD168723}, is also present in our sample. They observed short time scale variations and interpreted these as p-mode pulsations. The radial velocity observations by \citet{barban2004} have approximately the same amplitude as observed in the present survey. This indicates that the small radial velocity variations in the ``stable'' K giants are likely p-mode pulsations in the atmospheres of these stars. These pulsations are much more rapid than the typical time sampling of our observations, and thus appear as scatter in our data.

As the standard deviation is larger for all other stars in the sample (not considered in this paper), we can infer that essentially all K giants show radial velocity variations on the level of a few m/s. Furthermore since all stars with a standard deviation less than 20 m/s are found in the box in Figure~\ref{HRdiagram} (by definition), which ranges roughly to $B-V=1.2$, all stars redder than that show intrinsic variations larger than our threshold of 20 m/s. 

\subsection{Standard star sample}
The stars presented in this paper can serve as an addition to the standard star sample presented by \citet{udry1999a,udry1999b}. Only one star (\object{HIP19388}/\object{HD26162}) from the present survey is present in their sample. They obtained 252 observations for this star with a timespan of 6993 days and found a velocity dispersion of 0.3 km/s. This is the precision level of their observations and thus consistent with stability.
The 12 stars from the present survey with a flag set to K in Table~\ref{tab2} are also present among the 3967 candidate standards listed in Table 2 of \citet{kharchenko2004}. These candidate standards are selected based on the following criteria: no multiplicity or variability flag, standard errors of equatorial coordinates $\sigma < 40$ mas, standard errors of proper motions $\sigma_{pm} < 4$ mas/yr, standard errors of V magnitude $\sigma_{m_{V}} < 0.05$ mag and $B-V$ color $\sigma_{B-V} < 0.07$  mag, standard errors of radial velocity $\sigma_{RV} <  2$ km/s, and at least 4 radial velocity observations. The present observations could serve as a confirmation of their stability. 

The stars for which the flag in Table~\ref{tab2} is set to N are stars which do not have enough radial velocity observations in \citet{kharchenko2004} but do match all other criteria. The stars without a flag in Table~\ref{tab2} all have a photometric variability flag in the Tycho~1 catalog \citep{esa1997}, and are therefore not included in the \citet{kharchenko2004} candidate standard star catalog. For details see the main catalog, Table 1 of the same publication. However, from the present observations no evidence for variability in the radial velocity larger than 20 m/s is found.

\subsection{Reference stars}
The results presented in this paper provide a refined answer to the question that originally motivated our radial velocity survey, namely whether K giants are suited as reference stars for the Space Interferometry  Mission and other astrometric projects \citep[see] [ for details] {frink2001}. K giants are in principle good reference stars because they are intrinsically bright. Therefore sub-stellar companions do not disturb their photocenters much, neither by contributing light to the system, nor through their gravitational influence. (Note that for a desired apparent magnitude intrinsically brighter stars can be selected at a large distance, so that the angular displacement due to companions remains small.) However, any sample of ``anonymous'' rather distant K giants will contain a large fraction of binaries with stellar secondaries, which may lead to problems for astrometry. Our data demonstrate that binaries with a radial velocity  amplitude of a few tens of m/s can be identified readily with only a small number of high precision spectroscopic observations, provided that the giants chosen are not too red or too luminous. This reinforces the conclusion already drawn by \citet{frink2001} that K giants are indeed good astrometric reference stars, and validates the grid star strategy adopted by the SIM PlanetQuest project.

\subsection{Substellar companions and pulsations}
The radial velocity variations on the level of a few m/s are interesting for oscillation studies. With the present observations we show that they are observable in data with a precision of a few m/s. The time sampling of our observations is not suitable to obtain periods, but with campaigns taking multiple observations during each night this should be possible. Due to the fact that the oscillations appear on a level of a few m/s and with short periods it is also possible to search for (substellar) companions around these stars which can have larger radial velocity variations on longer timescales. 

\begin{acknowledgement}
SH wants to thank Ignas Snellen for a careful reading of this manuscript and very useful suggestions.
Furthermore, we would like to thank the entire staff at UCO/Lick Observatory for their untiring support.
\end{acknowledgement}
\bibliographystyle{aa}
\bibliography{bibstable}

\begin{thebibliography}{27}
\expandafter\ifx\csname natexlab\endcsname\relax\def\natexlab#1{#1}\fi

\bibitem[{{Allende Prieto} \& {Lambert}(1999)}]{allende1999}
{Allende Prieto}, C. \& {Lambert}, D.~L. 1999, \aap, 352, 555

\bibitem[{{Barban} {et~al.}(2004){Barban}, {de Ridder}, {Mazumdar}, {Carrier},
  {Eggenberger}, {de Ruyter}, {Vanautgaerden}, {Bouchy}, \&
  {Aerts}}]{barban2004}
{Barban}, C., {de Ridder}, J., {Mazumdar}, A., {et~al.} 2004, in ESA SP-559:
  SOHO 14 Helio- and Asteroseismology: Towards a Golden Future, 113

\bibitem[{{Barbier-Brossat} \& {Figon}(2000)}]{barbier2000}
{Barbier-Brossat}, M. \& {Figon}, P. 2000, \aaps, 142, 217

\bibitem[{{Butler} {et~al.}(1996){Butler}, {Marcy}, {Williams}, {McCarthy},
  {Dosanjh}, \& {Vogt}}]{butler1996}
{Butler}, R.~P., {Marcy}, G.~W., {Williams}, E., {et~al.} 1996, \pasp, 108, 500

\bibitem[{{Duflot} {et~al.}(1995){Duflot}, {Figon}, \&
  {Meyssonnier}}]{duflot1995}
{Duflot}, M., {Figon}, P., \& {Meyssonnier}, N. 1995, \aaps, 114, 269

\bibitem[{{ESA}(1997)}]{esa1997}
{ESA}. 1997, VizieR Online Data Catalog, 1239, 0

\bibitem[{Evans(1968)}]{evans1968}
Evans, D.~S. 1968, in Trans. IAU Vol XIII B, ed. L.~Perek (Reidel, Dordrecht),
  170

\bibitem[{{Famaey} {et~al.}(2005){Famaey}, {Jorissen}, {Luri}, {Mayor}, {Udry},
  {Dejonghe}, \& {Turon}}]{famaey2005}
{Famaey}, B., {Jorissen}, A., {Luri}, X., {et~al.} 2005, \aap, 430, 165

\bibitem[{{Frink} {et~al.}(2002){Frink}, {Mitchell}, {Quirrenbach}, {Fischer},
  {Marcy}, \& {Butler}}]{frink2002}
{Frink}, S., {Mitchell}, D.~S., {Quirrenbach}, A., {et~al.} 2002, \apj, 576,
  478

\bibitem[{{Frink} {et~al.}(2001){Frink}, {Quirrenbach}, {Fischer}, {R{\"o}ser},
  \& {Schilbach}}]{frink2001}
{Frink}, S., {Quirrenbach}, A., {Fischer}, D., {R{\"o}ser}, S., \& {Schilbach},
  E. 2001, \pasp, 113, 173

\bibitem[{{Haisch}(1999)}]{haisch1999}
{Haisch}, B.~M. 1999, in The many faces of the sun: a summary of the results
  from NASA's Solar Maximum Mission., 481

\bibitem[{{Hatzes} \& {Cochran}(1998)}]{hatzes1998}
{Hatzes}, A.~P. \& {Cochran}, W.~D. 1998, in ASP Conf. Ser. 154: Cool Stars,
  Stellar Systems, and the Sun, ed. R.~A. {Donahue} \& J.~A. {Bookbinder}, 311

\bibitem[{Heard(1968)}]{heard1968}
Heard, J.~F. 1968, in Trans. IAU Vol XIII B, ed. L.~Perek (Reidel, Dordrecht),
  169

\bibitem[{{Hekker} {et~al.}(2006){Hekker}, {Reffert}, \&
  {Quirrenbach}}]{hekker2006}
{Hekker}, S., {Reffert}, S., \& {Quirrenbach}, A. 2006, Communications in
  Asteroseismology, 147, 121

\bibitem[{{Henry} {et~al.}(2000){Henry}, {Fekel}, {Henry}, \&
  {Hall}}]{henry2000}
{Henry}, G.~W., {Fekel}, F.~C., {Henry}, S.~M., \& {Hall}, D.~S. 2000, \apjs,
  130, 201

\bibitem[{{Kharchenko} {et~al.}(2004){Kharchenko}, {Piskunov}, \&
  {Scholz}}]{kharchenko2004}
{Kharchenko}, N.~V., {Piskunov}, A.~E., \& {Scholz}, R.-D. 2004, Astronomische
  Nachrichten, 325, 439

\bibitem[{{Larson} {et~al.}(1999){Larson}, {Yang}, \& {Walker}}]{larson1999}
{Larson}, A.~M., {Yang}, S.~L.~S., \& {Walker}, G.~A.~H. 1999, in ASP Conf.
  Ser. 185: IAU Colloq. 170: Precise Stellar Radial Velocities, ed. J.~B.
  {Hearnshaw} \& C.~D. {Scarfe}, 193

\bibitem[{{Linsky} \& {Haisch}(1979)}]{linskyhaisch1979}
{Linsky}, J.~L. \& {Haisch}, B.~M. 1979, \apjl, 229, L27

\bibitem[{{Marcy} \& {Butler}(2000)}]{marbut2000}
{Marcy}, G.~W. \& {Butler}, R.~P. 2000, \pasp, 112, 137

\bibitem[{Mitchell {et~al.}(2006)Mitchell, Reffert, Quirrenbach, Fischer,
  Marcy, \& Butler}]{mitchell2006}
Mitchell, D.~S., Reffert, S., Quirrenbach, A., {et~al.} 2006, in prep.

\bibitem[{{Nidever} {et~al.}(2002){Nidever}, {Marcy}, {Butler}, {Fischer}, \&
  {Vogt}}]{nidever2002}
{Nidever}, D.~L., {Marcy}, G.~W., {Butler}, R.~P., {Fischer}, D.~A., \& {Vogt},
  S.~S. 2002, \apjs, 141, 503

\bibitem[{Pearce(1955)}]{pearce1955}
Pearce, J.~A. 1955, in Trans. IAU Vol IX B, ed. P.~Oosterhoff (Cambridge
  University Press, Cambridge), 441

\bibitem[{{Pepe} {et~al.}(2003){Pepe}, {Bouchy}, {Queloz}, \&
  {Mayor}}]{pepe2003}
{Pepe}, F., {Bouchy}, F., {Queloz}, D., \& {Mayor}, M. 2003, in ASP Conf. Ser.
  294: Scientific Frontiers in Research on Extrasolar Planets, 39

\bibitem[{{Queloz} {et~al.}(2001){Queloz}, {Mayor}, {Udry}, {Burnet},
  {Carrier}, {Eggenberger}, {Naef}, {Santos}, {Pepe}, {Rupprecht}, {Avila},
  {Baeza}, {Benz}, {Bertaux}, {Bouchy}, {Cavadore}, {Delabre}, {Eckert},
  {Fischer}, {Fleury}, {Gilliotte}, {Goyak}, {Guzman}, {Kohler}, {Lacroix},
  {Lizon}, {Megevand}, {Sivan}, {Sosnowska}, \& {Weilenmann}}]{queloz2001}
{Queloz}, D., {Mayor}, M., {Udry}, S., {et~al.} 2001, The Messenger, 105, 1

\bibitem[{{Reffert} {et~al.}(2006){Reffert}, {Hekker}, {Quirrenbach},
  {Fischer}, {Marcy}, \& {Butler}}]{reffert2006}
{Reffert}, S., {Hekker}, S., {Quirrenbach}, A., {et~al.} 2006, \aap, in
  preparation

\bibitem[{{Udry} {et~al.}(1999{\natexlab{a}}){Udry}, {Mayor}, {Maurice},
  {Andersen}, {Imbert}, {Lindgren}, {Mermilliod}, {Nordstr{\"o}m}, \&
  {Pr{\'e}vot}}]{udry1999a}
{Udry}, S., {Mayor}, M., {Maurice}, E., {et~al.} 1999{\natexlab{a}}, in ASP
  Conf. Ser. 185: IAU Colloq. 170: Precise Stellar Radial Velocities, 383

\bibitem[{{Udry} {et~al.}(1999{\natexlab{b}}){Udry}, {Mayor}, \&
  {Queloz}}]{udry1999b}
{Udry}, S., {Mayor}, M., \& {Queloz}, D. 1999{\natexlab{b}}, in ASP Conf. Ser.
  185: IAU Colloq. 170: Precise Stellar Radial Velocities, 367

\end{thebibliography}
\listofobjects
\end{document}